\documentclass[twocolumn,prl,aps,showpacs,superscriptaddress]{revtex4-1}
\usepackage{graphicx}
\usepackage[dvipdfm]{hyperref}
\usepackage{amsmath,gensymb}
\usepackage{epsfig}
\usepackage{natbib}
\newcommand{\ham}{\mathcal{H}}

\begin{document}
\title{Spin-Orbit Coupling Fluctuations as a Mechanism of Spin Decoherence}

\author{M. Martens}
\email{mmartens@fsu.edu}
\affiliation{Department of Physics, Florida State University, Tallahassee, Florida 32306, USA}
\affiliation{The National High Magnetic Field Laboratory, Tallahassee, Florida 32310, USA}

\author{G. Franco}
\affiliation{Department of Physics, Florida State University, Tallahassee, Florida 32306, USA}
\affiliation{The National High Magnetic Field Laboratory, Tallahassee, Florida 32310, USA}

\author{N.S. Dalal}
\affiliation{The National High Magnetic Field Laboratory, Tallahassee, Florida 32310, USA}
\affiliation{Department of Chemistry \& Biochemistry, Florida State University, Tallahassee, Florida 32306, USA}

\author{S. Bertaina}
\affiliation{Aix-Marseille Universit\'{e}, CNRS, IM2NP UMR7334, 13397 cedex 20, Marseille, France.}

\author{I. Chiorescu}
\email{ic@magnet.fsu.edu} 
\affiliation{Department of Physics, Florida State University, Tallahassee, Florida 32306, USA}
\affiliation{The National High Magnetic Field Laboratory, Tallahassee, Florida 32310, USA}

\date{\today}%
\begin{abstract}
We discuss a general framework to address spin decoherence resulting from fluctuations in a spin Hamiltonian. We performed a systematic study on spin decoherence in the compound K$_6$[V$_{15}$As$_6$O$_{42}$(D$_2$O)] $\cdot$ 8D$_2$O, using high-field Electron Spin Resonance (ESR). By analyzing the anisotropy of resonance linewidths as a function of orientation, temperature and field, we find that the spin-orbit term is a major decoherence source. The demonstrated mechanism can alter the lifetime of any spin qubit and we discuss how to mitigate it by sample design and field orientation. 
\end{abstract}
\maketitle
	
\section{Introduction}
In solid-state systems, interactions between electronic spins and their environment are the limiting factor of spin phase lifetime, or decoherence time. Important advances have been recently realized in demonstrating long-lived spin coherence via spin dilution \cite{Bertaina2009,Nellutla2007,Bertaina2007,Dutt2007,Bertaina_NatLett_2008,Ardavan2007} and isolating a spin in non-magnetic cages \cite{Morley2007}, for instance. The presence of a lattice can be felt by spins through orbital symmetries and spin-orbit coupling. An isolated free electron has a spin angular momentum associated with a $g$-factor $g_e = 2.00232$ but in general, spin-orbit coupling changes the $g$-factor by the admixture of excited orbital states \cite{Pryce_PPS_1950} into the ground state. In this Letter, we demonstrate that fluctuations in the spin-orbit interaction can be a significant source of spin decoherence.  We present a general theoretical framework to obtain noise spectrum. The method is applied to fluctuations of the long-range dipolar interactions and we observe how the spin-orbit term is modulating the induced decoherence. The model describes spin dilution and thermal excitations effects as well. Experimentally, we analyze shape and orientation anisotropy of ESR linewidths  of the molecular compound K$_6$[V$^{\text{IV}}_{15}$As$^{\text{III}}_6$O$_{42}$(D$_2$O)] $\cdot$ 8D$_2$O or V$_{15}$. This system has shown spin coherence at low temperatures~\cite{Dobrovitski2000,Bertaina_NatLett_2008} and interesting out of equilibrium spin dynamics due to phonon bottlenecking~\cite{Chiorescu_JMMM_2000,Chiorescu2000}. However, the details of the spin decoherence are still not fully understood. In the case of diluted or molecular spins, little evidence has been brought up to now on the role of spin-orbit coupling on spin coherence time. This study elucidates this decoherence mechanism and how to mitigate its effect.
	
\section{Fluctuations in Spin Hamiltonian}
\begin{figure}
		\centering
		\includegraphics[width=3.37 in]{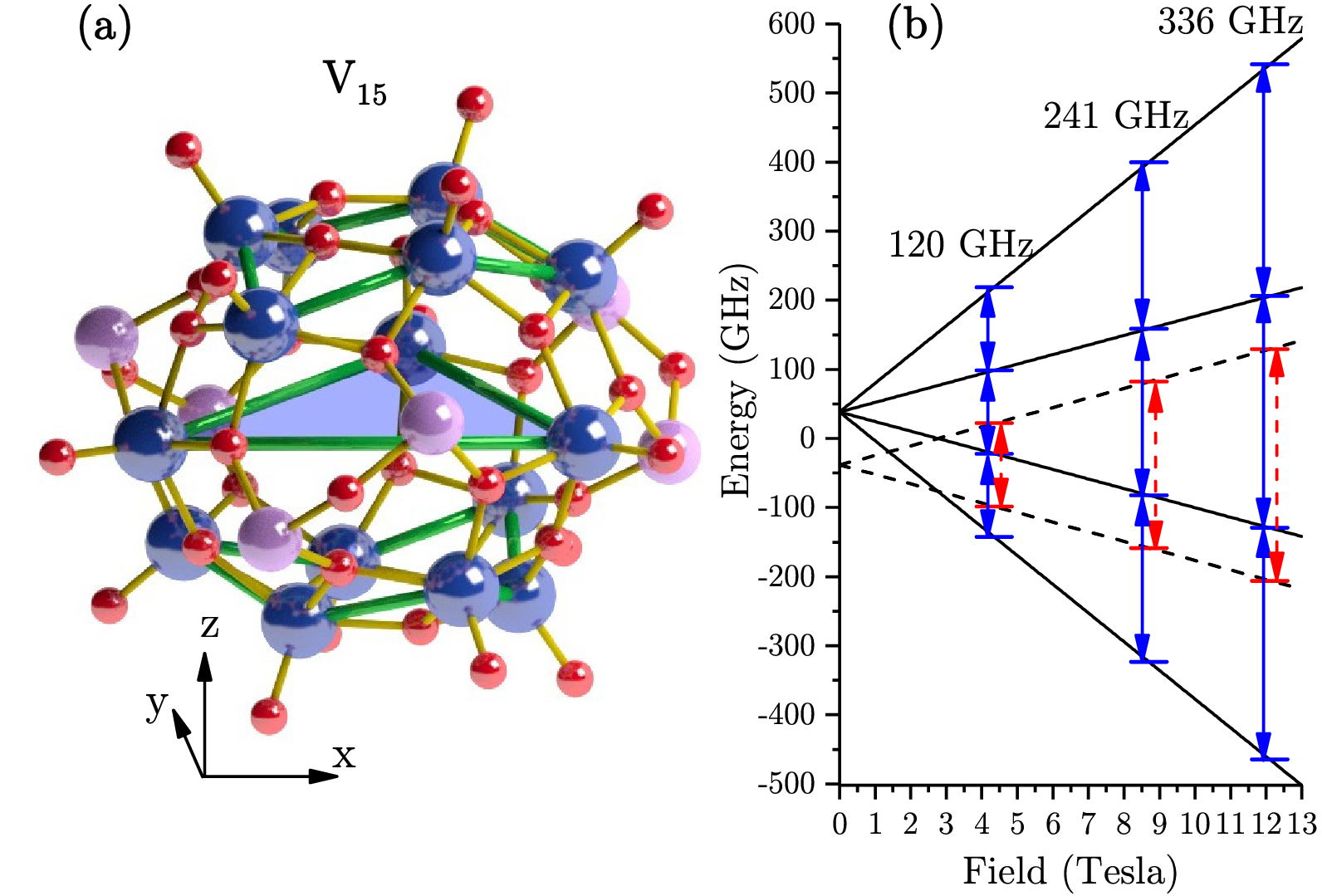}
		\caption{(Color online) (a) Ball-and-stick representation of V$_{15}$ (V ions in blue). The $x$ axis is along one side of the triangle while the $z$ axis is perpendicular to the triangle plane and represents the $c$ axis of the crystal unit cell. (b) Level diagram of the three spin model in field $||z$, with positions of the three experimental frequencies shown. Dashed lines show the $S=1/2$ doublets with the red dashed arrows indicating those transitions. Lines show the $S=3/2$ quartet with blue arrows indicating the transitions; the resonance fields are averaged in the first moment calculation of $B_{res}$ at a given frequency.}
		\label{fig:Mol_Elev}
	\end{figure}

The V$_{15}$ cluster anions form a lattice with trigonal symmetry containing two clusters per unit cell \cite{Mueller_AngChemie_1988}. Individual molecules have fifteen V$^{\text{IV}}$ $s=1/2$ ions arranged into three layers, two non-planar hexagons sandwiching a triangle (see Fig.~\ref{fig:Mol_Elev}a). Exchange couplings between the spins in the triangle and hexagons exceed 100~K~\cite{Gatteschi1991,Barra1992a} and at low temperatures this spin system can be modeled as a triangle of spins 1/2. The spin Hamiltonian is, as discussed in Supplemental Material\cite{SM_SOC} (SM) Section I:
	\begin{equation}\label{eq:Hstatic}
	\ham_{st} = \ham_0 + \ham_J + \ham_{DM}
	\end{equation}
where $\ham_0$ describes the Zeeman splitting in an external field $\vec{B}_0$, $\ham_J$ is the symmetric exchange term, and $\ham_{DM}$ is the anti-symmetric Dzyaloshinsky-Moriya (DM) term (see~\cite{Martens_PRB_2014} for a detailed formulation). $\ham_{st}$ eigenvalues are shown in Fig.~\ref{fig:Mol_Elev}(b) and are used to calculate resonant field positions $B_{res}$ of the ESR spectra through the method of first moments~\cite{Martens_PRB_2014}. As shown in Fig.~\ref{fig:Mol_Elev}(b), the ground state of the total molecular spin $\vec{S}$ is $S=3/2$ for large enough $\vec{B}_0$. In this case, dipolar interactions between total molecular spins in the crystal are described by:
	\begin{equation}
	\ham_{d} = \frac{3\mu_0}{8\pi} S^2 \mu_B^2 \sum_{p;q\neq p} g_p(\theta)g_q(\theta)\frac{ \left( 1 - 3 \cos^2 \phi_{p q} \right)}{d_{p q}^{3}} \label{eq:Hdz}
	\end{equation} 
 where $\mu_0$ is the vacuum permeability, $\mu_B$ is the Bohr magneton, $\theta$ is the angle between $\vec{B}_0$ and the $z$ axis ($z$ is $\perp$ to triangle plane and is also the symmetry $c$ axis of the molecule), $d_{pq}$ is the distance between two molecules located at sites $p$ and $q$, $g_{p,q}(\theta) = \left(g_a^2 \sin^2 \theta + g_c^2 \cos^2 \theta\right)^{1/2}$, $g_{c,a}$ are the $g$-tensor components parallel and perpendicular to the $z$ axis, $\phi_{pq}$ is the angle between $\vec{S}$ at site $p$ and $\vec{d}_{pq}$. Due to local fluctuations of the $g$-factor, as discussed below, $g_p$ and $g_q$ are distinct quantities.

The linewidth of ESR signals can be significantly affected by exchange interactions. In V$_{15}$ the intra-molecular couplings are large and the exchange narrowing effect~\cite{Kubo_JPSJ_1954} collapses the $\left(2I+1\right)^{15}$ resonances ($I=7/2$ for  $^{51}$V) into one and it also acts to average out fluctuations related to $\ham_{st}$. This leaves fluctuations in $\ham_{d}$ as being the major contributor to spin decoherence.
	
There are three possible sources of fluctuation in Eq.(\ref{eq:Hdz}), the first being the geometrical factor $ \left( 1 - 3 \cos^2 \phi_{p q} \right)d_{pq}^{-3} = R_{pq}(t)$ since both $d_{pq}$ and $\phi_{pq}$ can fluctuate (here, $t$ represents time). This case is described by \citet{Bloembergen_PR_1948} (Nuclear Magnetic Resonance case) and \citet{Kubo_JPSJ_1954} (ESR case). If $R_{pq}(t)$ fluctuates randomly, its correlation function decays exponentially $\langle R(t)R(0) \rangle = R^2 + r^2 \exp(-t/\tau_{dip})$ with a Fourier spectrum:
	\begin{equation}
		J_R(\nu) = \sqrt{\frac{2}{\pi}}r^2 \frac{\tau_{dip}}{1+4 \pi^2 \nu^2 \tau_{dip}^2}
		\label{eq:J}
	\end{equation}   
where $R$ is an average value of the geometric term $\sum_{p\neq q}R_{pq}$, $r$ is an average size of $R(t)$'s fluctuations and the correlation time $\tau_{dip}$ is a characteristic of the random motion. This result is described generally by \citet{Atherton_1973} and can be applied to any stationary random function that is independent of the time origin. The inverse square of the decoherence time $T_2$ is proportional to $\int J_R(\nu) d\nu$~\cite{Bloembergen_PR_1948,Kubo_JPSJ_1954}. Therefore, the decoherence rate depends directly on $r$: $1/T_2\propto r$.
	 	
Another fluctuation source comes from thermal excitations to different $S_Z$ states of $\vec{S}$, where $Z$ axis is $\parallel\vec{B}_0$, which defines the second moment of a resonance line~\cite{Pryce_PPS_1950_2,Kambe_PTP_1952,McMillan1960} (potential fluctuations between different spin states in low fields has been studied as well~\cite{Dobrovitski2000}): $\langle S_Z(t)^2S_Z(0)^2\rangle = S_Z^4+KU(T)\exp(-t/\tau_{s})$, where $\tau_s$ is the thermal correlation time and $KU(T)$ a term studied by Kambe and Usui \cite{Kambe_PTP_1952}. It is shown that the fluctuations Fourier spectrum is proportional to a temperature dependent factor:
	\begin{equation}
		KU(T)=<S_Z^2>_T-<S_Z>^2_T=S^2\frac{d}{dy}B_s(y)
		\label{eq:KU}
	\end{equation}
where $B_s(y)$ is the Brillouin function, $y=T_Z S/T$, $T_Z=hf_0/k_B$ ($f_0$ is the microwave excitation frequency), and $S=3/2$ is the total spin state. $KU(T)$ has thus a similar role to $r^2$ in Eq.~(\ref{eq:J}). This formulation is valid above the ordering temperature which is $\sim$0.01~K~\cite{Barbara_PTP_2002} for V$_{15}$.
	
The dipolar term $\ham_d$ serves as an excellent platform to study fluctuations of $g(\theta)$. Its value away from $g_e$ is due to the spin-orbit interaction and it is given by~\cite{Pryce_PPS_1950}:
	\begin{equation}
	\boldsymbol{g} = g_e \boldsymbol{I} - 2\lambda \boldsymbol{\Lambda}
	\end{equation} 
where $\boldsymbol{g}$ is the $g$-tensor (diagonal $[g_a,g_a,g_c]$ for V$_{15}$), $\boldsymbol{I}$ is the unit matrix, $\lambda$ is the spin-orbit coupling constant and $\boldsymbol{\Lambda}$ is a tensor defined in terms of the matrix elements of the orbital angular momentum $\mathbf{L}$. In general terms, $\Lambda$ is the coupling between the ground and excited orbitals divided by their energy separation. Relative fluctuations with an average size $\xi=\delta(\lambda\Lambda)/(\lambda\Lambda)$ (assumed isotropic) can be induced by crystal and molecular vibrations. In particular, Raman measurements on V$_{15}$ \cite{Zipse_PRB2005} discussed below, show a broad distribution of the vibration modes. Fluctuations of excited orbitals and thus of $\Lambda$ can generate broad virtual transitions since those orbitals are mixed with the ground orbital state. The resultant fluctuation in the $g$-factor can be written as:
	\begin{equation}
	\delta g(\theta) = \xi\left(g(\theta) - g_e\right).
	\label{eq:deltag}
	\end{equation}
	
Assuming $g(\theta)$ is a stationary function with small temporal random fluctuations and that magnetic and orbital fluctuations are uncorrelated in first approximation, the correlation function of a fluctuating $\ham_{d}(t)$ is:
	\begin{align}
	G_{d}(t)&= \langle \ham_{d}(t)\ham_{d}(0) \rangle\nonumber \\
	&=\alpha^2 \langle g(t)g(0)\rangle^2 \langle S_Z^2(t)S_Z^2(0)\rangle \langle R(t)R(0) \rangle
	\label{eq:deltaHdz}
	\end{align}
where $\langle g(t)g(0)\rangle = g(\theta)^2 + (\delta g(\theta))^2 \exp(-t/\tau_g)$, $\tau_{g}$ is the correlation time of $g$-factor fluctuations, and $\alpha = \frac{3\mu_0 \mu_B^2}{8\pi}$. A corresponding $J_{d}(\nu)$ gives the Fourier spectrum of the fluctuations, as in Eq.~\ref{eq:J}. $G_d(t)$ can be written as the sum of four terms (see SM Section II for details):  $G_0$ which is a constant , $G_g(t)\propto g(\theta)^4$, $G_\delta(t)$ which is temperature indepedent and $G_T(t)$ which is temperature dependent.
	
In absence of $g$-factor fluctuations, the resulting Fourier spectrum is defined only by $G_g(t)$ and for neglible $r$ (less important in solids at low temperatures) the term $G_g(t)\approx \alpha^2g(\theta)^4 R^2KU(T)e^{-\frac{t}{\tau_{s}}}$ is as in \cite{Kambe_PTP_1952}. A temperature dependence of the linewidth $\propto KU(T)$ is similar to observations done with Fe$_8$~\cite{Park_PRB_2002,Hill2002,Takahashi2009}, nitrogen-vacancy color centers in diamond~\cite{Takahashi2008} while other studies seem to confirm the proportionality to the $g$-factor~\cite{Graham2014,Takahashi2011}. If the $g$-value does fluctuate then all three terms $G_{g,\delta,T}$ represent sources of decoherence, with $G_\delta+G_T$ given by:
	\begin{align}
	&G_\delta(t)+G_T(t)\approx \alpha^2R^2\left[S^4+KU(T)\right]\times\nonumber\\
	&\times\left[ 2g(\theta)^2\delta g^2(\theta) e^{-\frac{t}{\tau_g}}+\delta g^4(\theta) e^{-\frac{2t}{\tau_g}}\right].
	\label{Gd}
\end{align}

Because $1/T_2^2$ is $\propto\int J_{d}(\nu) d\nu$, an important consequence is that one can combine different decoherence sources by summing their effect (each term $i$) as follows: $\frac{1}{T_2^2} \approx \sum_{i}\frac{1}{T_{2i}^2}$, similar to the well-known fact that the sum of uncorrelated variances is equal to the total variance. Additionally, the weight of each term in the sum depends on, or can be tuned with, the field angle $\theta$ through $g$ and $\delta g$. Here we show that for $V_{15}$, the anisotropy of the decoherence time is explained by fluctuations $\delta g$, as shown in Eq.~\ref{Gd}, amplified by spin thermal fluctuations $KU(T)$.
	
\section{Experimental Data}
Continuous-wave ESR measurements at 120, 241, and 336 GHz are performed using the quasioptical superheterodyne spectrometer at the National High Magnetic Field Laboratory~\cite{Morley_RSI_2008, vanTol_RSI_2005}, with a sweepable 12.5~T superconducting magnet (homogeneity of $10^{-5}$ over 1~cm$^3$). Sample temperature can be varied from room temperature down to 2.5 K. A single crystal of regular shape (as in \cite{Martens_PRB_2014}) of volume $\lesssim$ 0.1~mm$^3$ was positioned on a rotating stage allowing for continuous change of the angle $\theta$ between $\vec B_0$ and the $c$ axis of the molecule following the procedure described in~\cite{Martens_PRB_2014}. The homogeneity of the magnet compared to the size of the crystal allows us to ignore $\vec B_0$ as a source of broadening. The applied fields are above 4~T, past the crossing of the $S=1/2$ doublet and $S = 3/2$ quartet, such that the ground state of the system is in the $S = 3/2$ quartet (see Fig.~\ref{fig:Mol_Elev}(b)).
	\begin{figure}
	\centering
	\includegraphics[width=3.37 in]{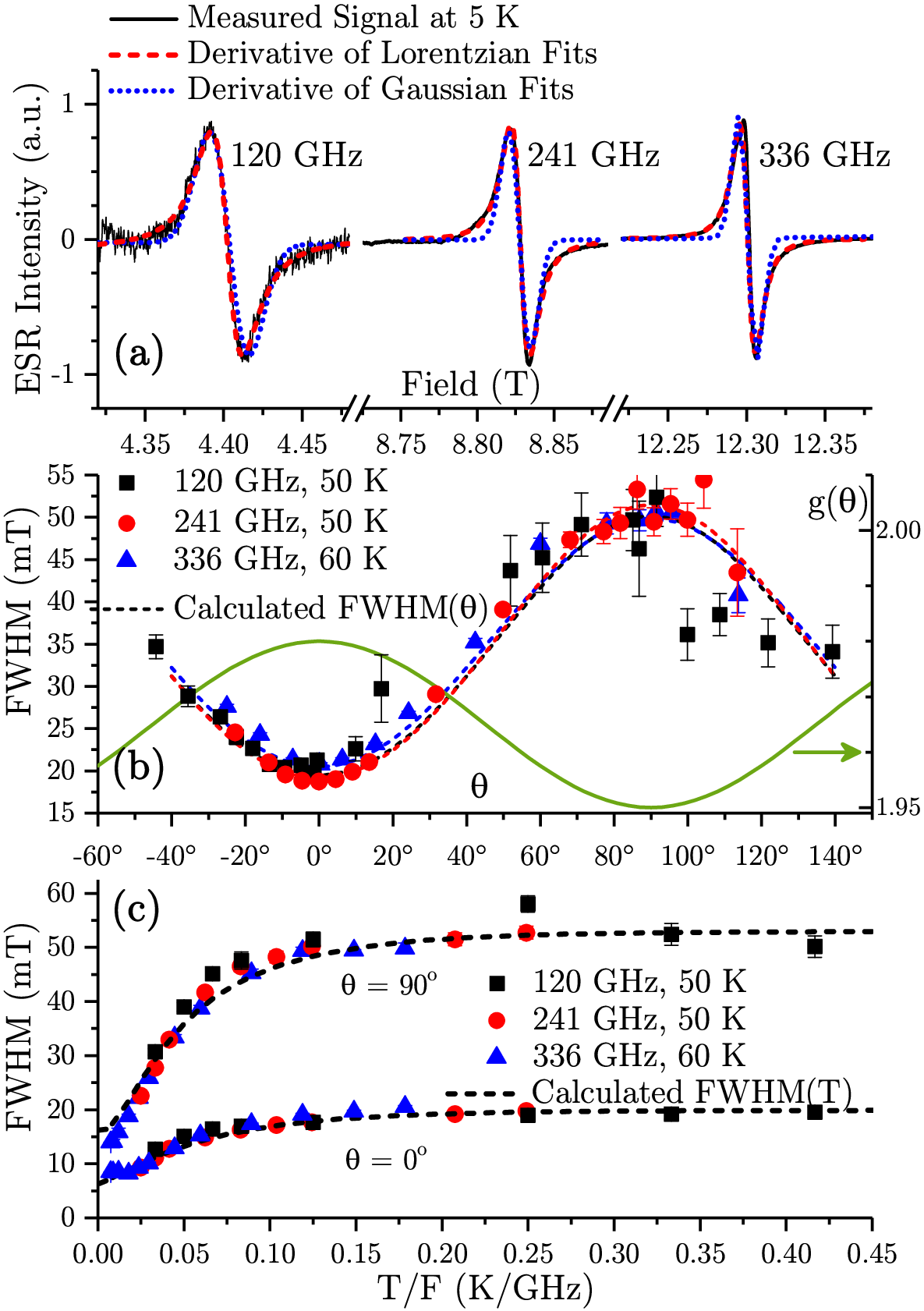}
	\caption{(Color online) (a) Typical measurements of the derivative of the absorption $\chi''$ at 120 GHz, 241 GHz and 336 GHz with derivative of Gaussian (blue dotted line) and Lorentzian fits (red dashed line). (b)  FWHM of Lorentzian fits as a function of field angle $\theta$ measured at three frequencies: 336 GHz (blue triangles), 241 GHz (red circles) and 120 GHz (black squares) The dashed lines are calculated widths as a function of $\theta$; the agreement shows the predicted correlation between decoherence rate and $g_e-g(\theta)$. In contrast, the green line (right axis) shows the opposite angular behavior of calculated $g(\theta)$, leading to $G_g\ll G_\delta+G_T$ (see text). (c) FWHM of Lorentzian fits vs temperature/frequency for the 3 studied frequencies. Dashed lines are calculated FWHM(T) for $\theta=0\degree$ and $90\degree$.}
	\label{fig:Tdep}
	\end{figure}
		
ESR spectra at temperatures $T=4-60$~K for $\vec B_0 \parallel$ and $\perp$ to $c$-axis ($\theta=0\degree ,90\degree$ respectively) show a Lorentzian (homogenous) lineshape.  Representative spectra with Lorentzian and Gaussian fits are shown in Fig.~\ref{fig:Tdep}(a) for comparison. The temperature dependence of the linewidth is shown in Fig.~\ref{fig:Tdep}(c) for three microwave frequencies $f_0$. Compared to measurements made at lower fields~\cite{Sakon_PhysB_2004}, where the ground state is in the $S=1/2$ doublet, the linewidths are $\sim10$ times narrower. Plotted is the full width at half maximum (FWHM) of the Lorentzian fits vs $T/f_0$ to underline that the temperature dependent mechanism of decoherence in the system qualitatively follows the temperature behavior predicted by $KU(T)$  plus a temperature independent contribution, essential in the low $T$ limit [Eq.~(\ref{Gd})]. Note the absence of linewidth increase with frequency (or field) which excludes a static distribution of the $g$-factor. Additionally, there is no hyperfine structure visible in the spectra (exchange narrowing) since the $3d$ electrons of V interact with the nuclei of several other V ions due to the large exchange couplings ($\sim 10^2$~K) within the molecule. Due to these properties of the measured line width and shape, we can estimate $T_2$ to be the inverse of the FWHM.
	
There are two distinct curves in Fig.~\ref{fig:Tdep}(c), dependent on the orientation of $\vec{B}_0$. To probe this orientation dependence, the linewidth is measured as a function of $\theta$, see Fig.~\ref{fig:Tdep}(b). The narrowest linewidth occurs when $\theta = 0^{\circ}$ ($\vec{B}_0 \parallel c$ axis) while the largest occurs at $\theta = 90^{\circ}$ ($\vec{B}_0$ in the triangle plane). This implies more decoherence the further $g(\theta)$ gets from $g_e$ since $g(0^{\circ})=g_c \approx 1.98$ and  $g(90^{\circ})=g_a \approx 1.95$. The fact that the width is largest (smallest) when $g(\theta)$ is minimum (maximum) rules out exchange narrowing being the cause of this anisotropy since it would require a linewidth $\propto(1+\cos^2\theta)$~\cite{Kubo_JPSJ_1954}, as in the case of CsCuCl$_3$~\cite{Tanaka_JPSJ_1981}. To rule out an angular effect of the dipolar field distribution, we measured  the FWHM($\theta$) on a sample of irregular shape at 240 GHz and 60 K. Although the shape-dependent coefficients $R$ and $r$ must be different, the same behavior is observed as in Fig.~\ref{fig:Tdep}b (see SM Section III B for details). We can thus focus on the terms $G_{g,\delta,T}$ as source of fluctuations.

However, the term $G_g(t)$ is $\propto g(\theta)^4$, in clear contrast with the observation that the FWHM and $g(\theta)$ have opposite angular dependences (see Fig.~\ref{fig:Tdep}b and also SM Section IIIA). This is the essential property of the V$_{15}$ compound, which makes it particularly suitable to study the effect of spin-orbit fluctuations. Therefore, this opposite angular behavior provides evidence that $\delta g(\theta)\neq 0$ and the terms $G_{\delta,T}(t)$ must be considered while $G_g(t)$ can be discarded. One can argue that geometrical fluctuations in solids at low temperatures are very small ($r\ll R$) and lattice fluctuations are mostly influencing the relaxation time $T_1$ ($\tau_s\gg\tau_g$) making $G_g\approx$ constant at the time scale of the decoherence time.

Since $G_d(t)\approx G_\delta(t)+G_T(t)$ and $1/T_2^2 \propto \int J_{d}(\nu)$ ~\cite{Bloembergen_PR_1948,Kubo_JPSJ_1954} the linewidth square  can be modeled by the following fit function (see SM Section II for details):
	\begin{align}\label{eq:angularfit}
	\Delta^2=&\left[S^4+KU(T)\right]\times\nonumber\\
	&\times \left[2a^2g(\theta)^2(g(\theta)-g_e)^2 +A^2 (g(\theta)-g_e)^4\right] 
	\end{align}
where $A$ and $a$ are fit parameters. The procedure is detailed in SM Section IV; it allows to calculate the angular dependence FWHM$(\theta)$ by using only two data points, $\Delta(0^\circ)$ and $\Delta(90^\circ)$, as shown in Fig.~\ref{fig:Tdep}(b) (dashed lines).
	 
To analyze the temperature dependence of the linewidth shown in Fig.~\ref{fig:Tdep}(c), we solve for $A$ and $a$ at all available temperatures and frequencies (see SM Section IV B for details). Above 10-20~K, the values stabilize at $A\sim$100~GHz and $a\sim$3.2~GHz. At lower temperatures, the values decrease by almost half, indicating a small decrease in $\xi$ and/or a slowing down in the fluctuations time $\tau_g$. These temperature trends $A(T)$ and $a(T)$ are estimated by an exponential saturation (see SM Fig.~4 for details), with decay constants of 3.6~K and 11 K for $A(T)$ and $a(T)$ respectively. With no other adjustments, the calculated linewidth is in very good agreement with the experimental data, as shown with dashed line in Fig.\ref{fig:Tdep}(c). On the low end of $T/f_0$ one observe a residual value of the linewidth, which includes the effects of other decoherence sources (such as the nuclear spin bath~\cite{Prokofev1998}), although it can be well described by Eq.~\ref{eq:angularfit}.

The outcome of the fit procedure can be used to estimate the size of spin-orbit fluctuations (see SM Section IV C) leading to an order of magnitude for $\xi\sim10^{-2}$. This corresponds to a fluctuation $\delta g/g\sim 10^{-4}$, too small to result in directly measurable fluctuations of the Zeeman splitting. Note that for V$_{15}$, a large spin-orbit fluctuation is supported by previous Raman measurements \cite{Zipse_PRB2005} showing a very broad signal in the region of $\sim500$~cm$^{-1}$ corresponding to vibrations of oxygen bridges between V ions. The observed broad distribution of the modes can induce very fast virtual transitions to excited coupled states and, as a consequence, spin decoherence.

\section{Conclusion}

Our study provides insight on how to mitigate the effects of spin-orbit fluctuations. It is evident from Eqs.~(\ref{eq:deltag}) and (\ref{eq:deltaHdz}) that the $g$-tensor should be as close as possible to $g_e$. In molecular compounds this can be achieved by engineering the ligands type since local symmetry affects the diagonal values of the  $g$-tensor of a magnetic ion. Aside from material design by chemical methods, $J_{d}(\nu)$ can be minimized by applying the magnetic field at a specific angle $\theta$. For V$_{15}$, this would be $\theta=0$ for which the decoherence time reaches several nanoseconds. This time can reach $\sim400$~ns by reducing $R$ in $J_{d}(\nu)$ via dilution in liquid state, thus allowing the observation of Rabi oscillations and spin-echoes~\cite{Bertaina_NatLett_2008}. The methodology presented here can be important for the diluted spin systems as well, since long range interactions are still present and can carry modulations due to $g$-factor fluctuations. Potential examples are transition metals such as Cr$^{5+}$:K$_3$NbO$_8$\cite{Nellutla2007}) or some lanthanide monomers doped into insulating lattice such as Hf$^{3+}$:LuPO$_4$\cite{Abraham_JCP1985} or La$^{2+}$:CaF$_2$\cite{Hayes_PPS1963}. The results extend to any solid-state system where spin-orbit coupling leads to quantum effects, independent of system dimensionality.

\section*{Acknowledgements}

We wish to acknowledge David Zipse and Vasanth Ramachandran for their  help in growing  V$_{15}$  crystals. This work was supported by NSF Grant No. DMR-1206267 and CNRS-PICS CoDyLow. The NHMFL is supported by Cooperative Agreement Grant No. DMR-1157490 and the state of Florida. The experiments were performed at the EPR facilities of the NHMFL, and we thank dr. Johan van Tol for his support.

\bibliographystyle{apsrev4-1} 
\bibliography{refs}

\newpage

\section{SUPPLEMENTAL MATERIAL: \\ Spin-Orbit Coupling Fluctuations as a Mechanism of Spin Decoherence}
\section{Single Molecule Spin Hamiltonian}

The molecular compound K$_6$[V$^{\text{IV}}_{15}$As$^{\text{III}}_6$O$_{42}$(D$_2$O)] $\cdot$ 8D$_2$O or V$_{15}$ contains fifteen V$^{+4}$ spins, each with a spins 1/2, shown in blue in Figure~1(a) of the article. The spins are coupled by large antiferromagnetic (AF) couplings in exces of 100~K~\cite{Gatteschi1991,Barra1992a}. At low temperatures, the molecule can be modeled as an effective frustrated triangle of spins 1/2. The effective antiferromagnetic coupling is $\approx2.51$~K as estimated by magnetic measurements at low temperature~\cite{Barbara_PTP_2002}. The generalized three spin Hamiltonian $\ham_{st}$ is given in the article as:
\begin{equation}\label{eq:Hstatic}
\ham_{st} = \ham_0 + \ham_J + \ham_{DM}.
\end{equation}
where $\ham_0$ describes the Zeeman splitting in an external field $\vec{B}_0$, $\ham_J$ is the symmetric exchange term, and $\ham_{DM}$ is the anti-symmetric Dzyaloshinsky-Moriya (DM) term. In the study presented here, the focus lies on the fluctuations of the dipolar term. However, for completness of the study, a summary of the properties of the static term $\ham_{st}$ is presented in this section following our previous study, Ref.~[\onlinecite{Martens_PRB_2014}].

The Zeeman term is given by:
\begin{equation}
\ham_0=g_a\mu_B\vec B_\triangle \vec S_\triangle + g_c\mu_B\vec B_z \vec S_z
\end{equation}
where $g_{a,c}$ are the values of the $g$-factor in the triangle's plane and along the molecule's $c$ axis ($z$ axis in Fig.~1a) respectively; $\vec B_{\triangle,z}$ are the planar and vertical vector components of the external magnetic field $\vec{B}_0$, with $B_\triangle=B_0\sin\theta$ and $B_z=B_0\cos\theta$, respectively; $\theta$ is the angle between $\vec{B}_0$ and the $z$-axis;  $\vec S_{\triangle}=\vec S_x+\vec S_y= \sum_{i=1,2,3}(\vec S^{(i)}_x+\vec S^{(i)}_y) $, and $\vec S_{z}=\sum_{i=1,2,3}\vec S^{(i)}_z$.

\begin{figure}
	\includegraphics[width=\columnwidth]{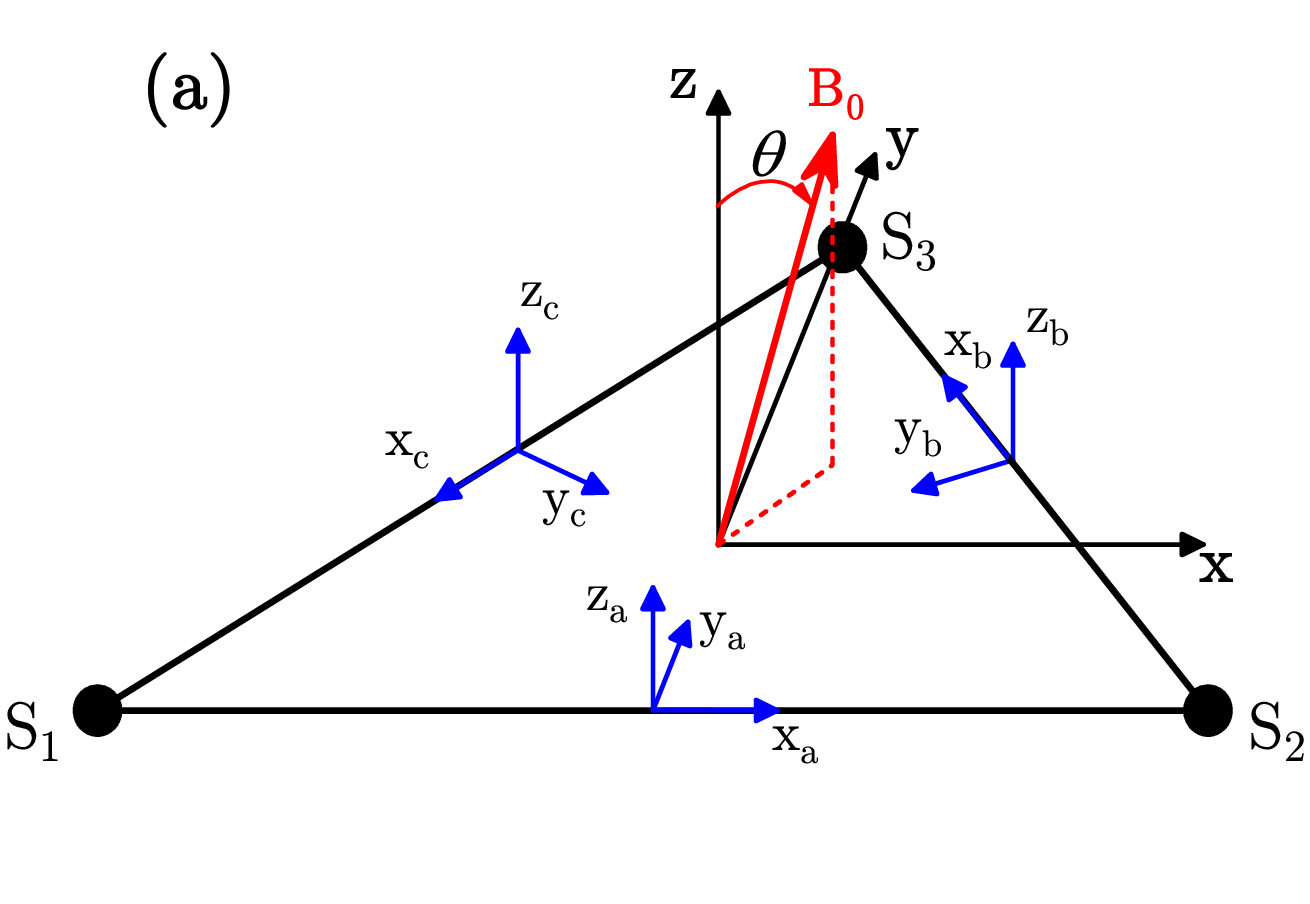}
	\includegraphics[width=\columnwidth]{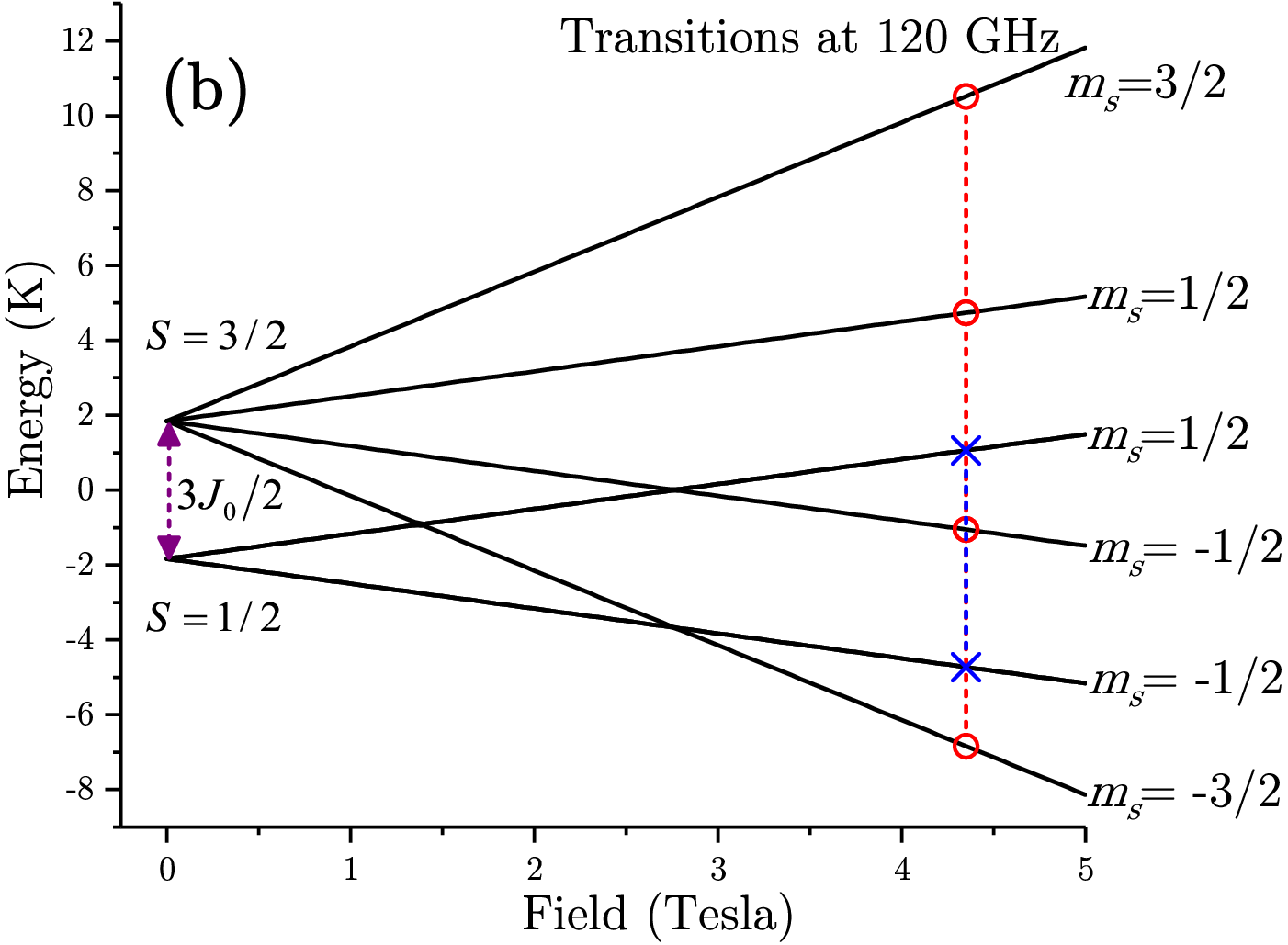}
	\caption{(color online) (a) Diagram showing the local $a,b,c$ axes defining the DM vectors in the three spin model. The spins are located in the triangle corners. The applied field $\vec{B}_0$ makes an angle $\theta$ with the $z$-axis $\perp$ to the triangle plane. (b) Energy eigenstates of the three spin model with positions of allowed transitions at 120 GHz shown. Red dashed lines between red open circles indicate transitions inside the $S=3/2$ quartet while the blue dashed line between the blue crosses indicate the four allowed transitions within the degenerate $S=1/2$ doublets. The zero-field splitting between the quartet and doublets is $3J_0/2$.}
	\label{fig:V15Elev}
\end{figure}

The exchange term is given by:
\begin{equation}
\ham_J=\sum_{<i,j>}[J_z^{(ij)} S_z^{(i)} S_z^{(j)} + J_t^{(ij)}(S_x^{(i)}S_x^{(j)}
+S_y^{(i)}S_y^{(j)})]
\end{equation}
and the DM term by:
\begin{equation}
\ham_{DM}=\sum_{<i,j>}\vec D^{(ij)}(\vec S^{(i)}\times\vec S^{(j)})
\end{equation}
where one sums over the $(i,j)$ pairs $(1,2),(2,3),(3,1)$ of the triangle spins $\vec S^{(1),(2),(3)}$. The $\ham_J$ term describes the effective antiferromagnetic coupling between the triangle's corners which, in principle, can be anisotropic ($J_z\ne J_t$) and differ between spin-pairs in the triangle ($J_z^{(12)} \neq J_z^{(23)}$ for example). In the $\ham_{DM}$ term, the individual vectors $\vec D^{(ij)} = \vec D_x^{(ij)} + \vec D_y^{(ij)}+ \vec D_z^{(ij)}$ correspond to the interaction between $S^{(i)}$ and $S^{(j)}$, and are related to the local frames ($a$, $b$, and $c$) shown in \autoref{fig:V15Elev}(a). It is assumed that $D_{x_a}^{(12)} = D_{x_b}^{(23)} = D_{x_c}^{(13)} = D_l$, $D_{y_a}^{(12)} = D_{y_b}^{(23)} = D_{y_c}^{(13)} = D_t$, and $D_{z_a}^{(12)} = D_{z_b}^{(23)} = D_{z_c}^{(13)} = D_z$.

Diagonalizing $\ham_{st}$ yields an energy structure that is made up of an excited quartet $S=3/2$ and two ground doublets $S=1/2$ where the zero-field splitting between the quartet and doublets is determined by the value of the antiferromagnetic coupling. This is shown in \autoref{fig:V15Elev}(b) for the case where there is no DM interaction, $J_z^{(ij)} = J_t^{(ij)} = J_0 = -2.45$~K for all $(ij)$, $g_a=1.95$ and $g_c=1.98$. In this simple case, the ground doublets are degenerate and are separated from the quartet at zero field by $3 J_0 /2$. The crossover of the ground state from the doublet to the quartet is known to happen at 2.8 T from measurements of the magnetization of a single V$_{15}$ crystal,\cite{Barbara_PTP_2002} placing a constraint on the values of $J_z^{(ij)}$ and $J_t^{(ij)}$. 

The parameters of $\ham_{st}$ can be identified by means of high-field spin resonance spectroscopy as described in a previous study\cite{Martens_PRB_2014}. The excitation frequency of 120~GHz corresponds to fields upwards of 4~T, where the quartet represents the ground state. The allowed transitions are shown in \autoref{fig:V15Elev}(b) with dashed vertical lines, where the blue line between the crosses represents the four allowed transitions within the degenerate doublets. The field positions of these transitions depend on the values of all $\ham_{st}$ terms and each transition has a certain effect on the overall field location of the resonance signal. By using numerical diagonalization and the first moment method to average the transitions, information on the spin Hamiltonian parameters can be extracted. Such approach is very general and can be applied to solve a non-linear relationship between magnetic field and resonance frequency. The conclusion of the study can be summarized as follows. Any isotropy in the AF exchange coupling $J$ or generated by the DM term is not sufficiently large to create a spectroscopic signature. The data together with the numerical calculations of the transition probabilities indicate that the doublet transitions, shown in blue arrows in \autoref{fig:V15Elev}(b), don't have any noticeable influence on high-field spin resonance experiments.

\section{Spectral density of dipolar fluctuations}

The interaction between V$_{15}$ neighboring molecules is of dipolar nature and described by Eq.~(2) of the main article:
\begin{equation}
\ham_{d} = \frac{3\mu_0}{8\pi} S^2 \mu_B^2 \sum_{p;q\neq p} g_p(\theta)g_q(\theta)\frac{ \left( 1 - 3 \cos^2 \phi_{p q} \right)}{d_{p q}^{3}} 
\label{eq:Hdz}
\end{equation} 
with $\mu_0$ the vacuum permeability, $\mu_B$ the Bohr magneton, $d_{pq}$ the distance between two molecules located at sites $p$ and $q$, $g_{p,q}(\theta) = \left(g_a^2 \sin^2 \theta + g_c^2 \cos^2 \theta\right)^{1/2}$, $\phi_{pq}$ the angle between $\vec{S}$ at site $p$ and $\vec{d}_{pq}$. Random fluctuations in $\ham_d$ generate the observed linewidth and thus limit the spin coherence lifetime. 

To find the spectral density of dipolar fluctuations one has to analyse the time dependence of the correlation function, defined in the main article as $G_d(t)=\langle \ham_d(t)\ham_d(0)\rangle=\alpha^2 \langle g(t)g(0)\rangle^2 \langle S^2(t)S^2(0)\rangle \langle R(t)R(0) \rangle$:

\begin{align}
G_{d}(t)&= \alpha^2\left[g(\theta)^2 + \delta g^2(\theta) e^{-\frac{t}{\tau_g}}\right]^2 \nonumber \\
&\phantom{=}\times\left[S^4+KU(T)e^{-\frac{t}{\tau_{s}}}\right]\left[R^2 + r^2 e^{-\frac{t}{\tau_{dip}}}\right] \nonumber \\
&=\alpha^2 \left[g(\theta)^4 + 2g(\theta)^2\delta g^2(\theta) e^{-\frac{t}{\tau_g}}+\delta g^4(\theta) e^{-\frac{2t}{\tau_g}}\right]\nonumber \\
&\phantom{=}\times\left[S^4+KU(T)e^{-\frac{t}{\tau_{s}}}\right] \left[R^2 + r^2 e^{-\frac{t}{\tau_{dip}}}\right] 
\end{align}
where all notations are defined in the main article. Thus, $G_d(t)$ is a twelve term sum which can be split as follows:
\begin{align}
G_0&=\alpha^2g(\theta)^4S^4R^2,\nonumber \\
G_g(t)&=\alpha^2g(\theta)^4R^2KU(T)e^{-\frac{t}{\tau_{s}}}+\nonumber \\
&\phantom{=} +\alpha^2g(\theta)^4r^2 e^{-\frac{t}{\tau_{dip}}}\left(S^4+KU(T)e^{-\frac{t}{\tau_{s}}}\right) ,\nonumber \\
G_\delta(t)&=\alpha^2S^4\left[2g(\theta)^2\delta g^2(\theta) e^{-\frac{t}{\tau_g}}+\delta g^4(\theta) e^{-\frac{2t}{\tau_g}}\right] \nonumber \\
&\phantom{=} \times\left[R^2+r^2 e^{-\frac{t}{\tau_{dip}}}\right],\nonumber \\
G_{T}(t)&=\alpha^2KU(T)e^{-\frac{t}{\tau_{s}}}\left[2g(\theta)^2\delta g^2(\theta) e^{-\frac{t}{\tau_g}}+\delta g^4(\theta) e^{-\frac{2t}{\tau_g}}\right] \nonumber \\
&\phantom{=} \times \left[R^2+r^2e^{-\frac{t}{\tau_{dip}}}\right].
\end{align}

As discussed in the main article, one can follow Atherton\cite{Atherton_1973} to obtain the Fourier spectrum corresponding to each term above. The first term, $G_0$, is time indepedent and thus will not count. Also, given the experimental data for the particular case of V$_{15}$, the term $G_g$ must not be significant. This is a result of the opposite angular dependence between measured $g(\theta)$ and signal linewidth, an analysis given in Section~\ref{g4}. This condition imposes that $r^2$ is very small, meaning that crystal lattice fluctuations at low temperatures are not of significance for the decoherence rate. Such process is significant for the relaxation time $T_1$, related to $\tau_s$ and usually much larger than the decoherence time. Consequently, $\tau_s \gg \tau_g, \tau_{dip}$ and therefore the first term in $G_g(t)$ is almost constant, at the scale of the decoherence time. Such condition is valid in a large range of temperatures and fields, leaving only two dominant terms, $G_\delta$ and $G_T$. 

The essential difference between $G_\delta$ and $G_T$ is that the former is temperature independent while the later describes the measured temperature dependence (see Fig.~2b). At the low end of the temperature range, one achieve a minimization of the continuous wave linewidth. This can be attributed to the temperature independent term $G_\delta$ plus any other decoherence sources, such as the nuclear spin bath~\cite{Prokofev1998}.

Consequently, the correlation function can be written as:
\begin{align}
G_d(t)\approx G_\delta(t)+G_T(t)\approx\alpha^2\left[S^4+KU(T)\right]R^2  \nonumber \\
\times\left[2g(\theta)^2\delta g^2(\theta) e^{-\frac{t}{\tau_g}}+\delta g^4(\theta) e^{-\frac{2t}{\tau_g}}\right],
\end{align}
and the resulting Fourier spectrum is:
\begin{align}
J_d(\nu)&=\sqrt{\frac{2}{\pi}}\alpha^2\left[S^4+KU(T)\right]R^2 \nonumber \\
\times& \left[2g(\theta)^2\delta g^2(\theta) \frac{\tau_g}{1+4\pi^2\tau_g^2\nu^2} + \delta g^4(\theta)\frac{\tau_{g}/2}{1+\pi^2\tau_{g}^2 \nu^2} \right].
\end{align}

Since the integral of $J_d(\nu)$ represents the square of the decoherence rate, $1/T_2^2=\int J_{d}(\nu) d\nu$, one can conclude that the appropiate fit function for the spectroscopy linewidth is given by:
\begin{align}
\Delta^2&= \sqrt{\frac{2}{\pi}}\alpha^2\left[S^4+KU(T)\right]R^2 \nonumber \\
&\phantom{=} \times \left[2g(\theta)^2\xi^2(g(\theta)-g_e)^2\int \frac{\tau_g}{1+4\pi^2\tau_g^2\nu^2}d\nu \right]+ \nonumber \\
&\phantom{=}+ \sqrt{\frac{2}{\pi}}\alpha^2\left[S^4+KU(T)\right]R^2 \nonumber \\
&\phantom{=} \times\left[\xi^4(g(\theta)-g_e)^4\int \frac{\tau_{g}/2}{1+\pi^2\tau_{g}^2 \nu^2} d\nu \right],
\label{Deltaxi}
\end{align}
where $\xi=\delta g (\theta)/(g(\theta)-g_e)$ and $g_e=2.00232$. This leads to a fit function of the experimental linewidth $\Delta$:

\begin{align}
\Delta^2&=\left[S^4+KU(T)\right] \nonumber \\
&\phantom{=} \times\left[2a^2g(\theta)^2(g(\theta)-g_e)^2 + A^2 (g(\theta)-g_e)^4\right],
\label{Delta}
\end{align}
where $a$ and $A$ are fit parameters.

\section{Angular dependence of the linewidth/spectral density}

\subsection{The $g^4$ term is negligible}\label{g4}
When discussing the correlation function $G_d(t)$ above, an approximation was made, namely that the $G_g(t)$ term must not be dominating because it contains the factor $g(\theta)^4$. The V$_{15}$ molecular compound has a very particular propriety, namely the behavior of the linewidth when the angle $\theta$ is changed continously. Our measurements (see Fig.~2b of the main article) show an opposite angular dependence between linewidth and $g(\theta)$. At the extreme points $\theta=0^{\circ}$ and $90^{\circ}$ a small/large linewidth is obtained when the $g$-factor is the largest/smallest, with  $g(0^{\circ})=g_c \approx 1.98$ and  $g(90^{\circ})=g_a \approx 1.95$. This is somewhat counter-intuitive since a large $g$-factor means a larger magnetic moment and thus potentially larger dipolar fluctuations.

This unusual behavior can not be described by the term $g^4(\theta)$ which corresponds to a traditional interpretation of dipolar fluctuations: the larger the value of the dipoles ($\propto g,S$) the larger the value of $\ham_d$ fluctuations. Instead, the linewidth is in sync with the angular dependence of $\delta(\theta)$, that is the size of the shift of the $g$-factor away from the free electron value $g_e$. This shift of $g$ is caused solely by spin-orbit coupling \cite{Pryce_PPS_1950}: $\boldsymbol{g} = g_e \boldsymbol{I} - 2\lambda \boldsymbol{\Lambda}$ where $\boldsymbol{g}$ is the $g$-tensor (diagonal $[g_a,g_a,g_c]$ for V$_{15}$), $\boldsymbol{I}$ is the unit matrix, $\lambda$ is the spin-orbit coupling constant and $\boldsymbol{\Lambda}$ is a tensor defined in terms of the matrix elements of the orbital angular momentum $\mathbf{L}$:
\begin{equation}
\Lambda_{ij}=\sum_{n\neq 0}\frac{\langle 0|L_i | n \rangle \langle n| L_j | 0 \rangle}{\left( E_n - E_0 \right)}
\end{equation} 
with $i,j=x,y,z$. The spin-orbit coupling to excited orbitals is defining the value of the $g$-factor and vibrational modes will induce fluctuations in the term $\lambda\Lambda$. As a consequence to the opposite angular behavior of the linewidth and the $g$-factor, we conclude that $\delta g$ fluctuations with a characteristic time $\tau_g$ modulate the dipolar fluctuations. This makes the terms in $\delta g^2$ and $\delta g^4$, rather than $g^4$, to dominate the spectral density $J_d(\nu)$ shown above.

\subsection{Dipolar Hamiltonian: the geometric factor}

\begin{figure}
		\includegraphics[width=\columnwidth]{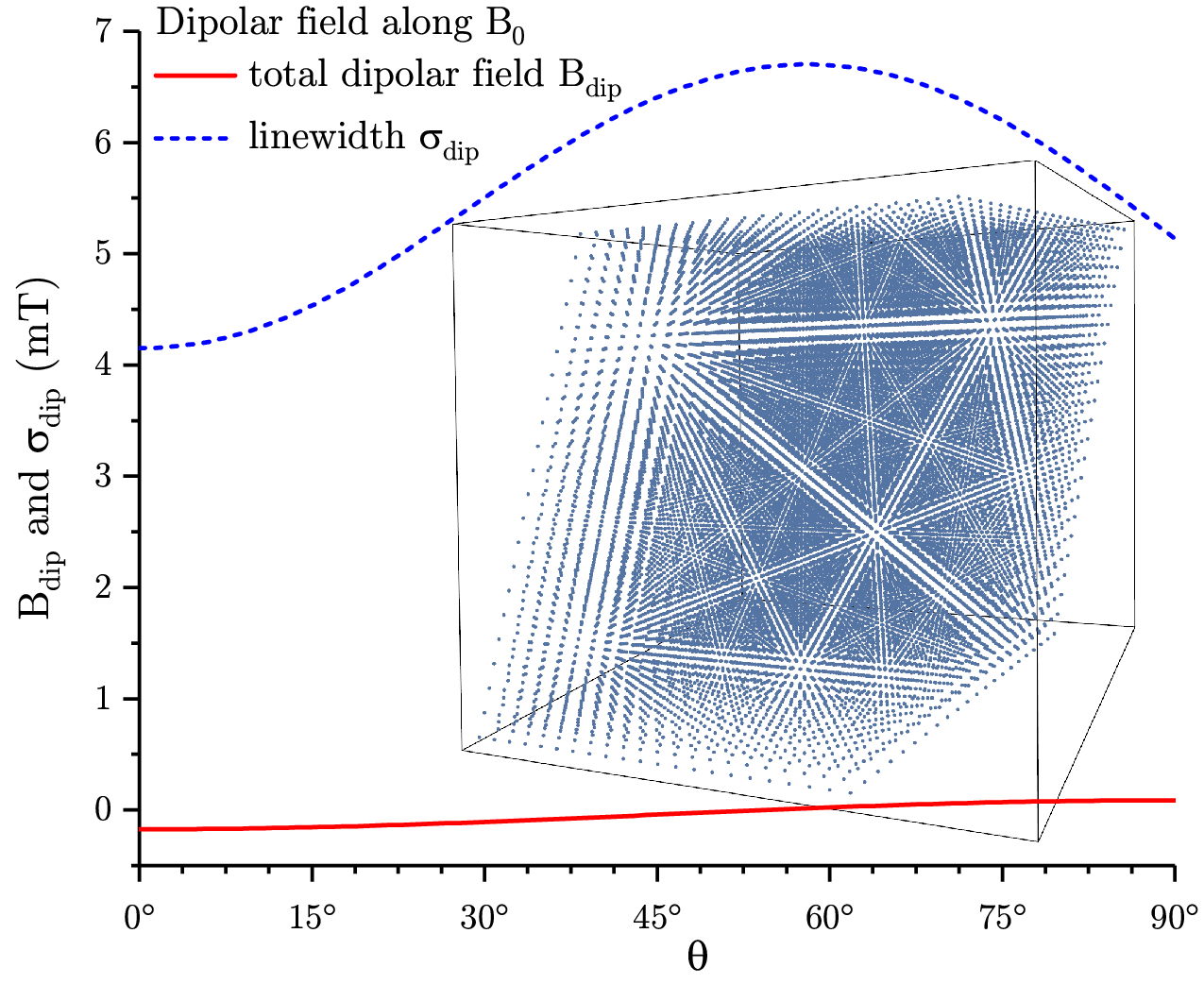}
		\caption{(color online) Dipolar field (continous line) and $\sigma_{dip}$ (dashed line) as a function of $\theta$ calculated in the middle of an ensemble of 10648 molecular spins, shown in the insert. The $B_{dip}$ has a typical $\propto1-3\cos^2\theta$ behavior.}
		\label{figSI_dipole}
\end{figure}
\begin{figure}
		\includegraphics[width=\columnwidth]{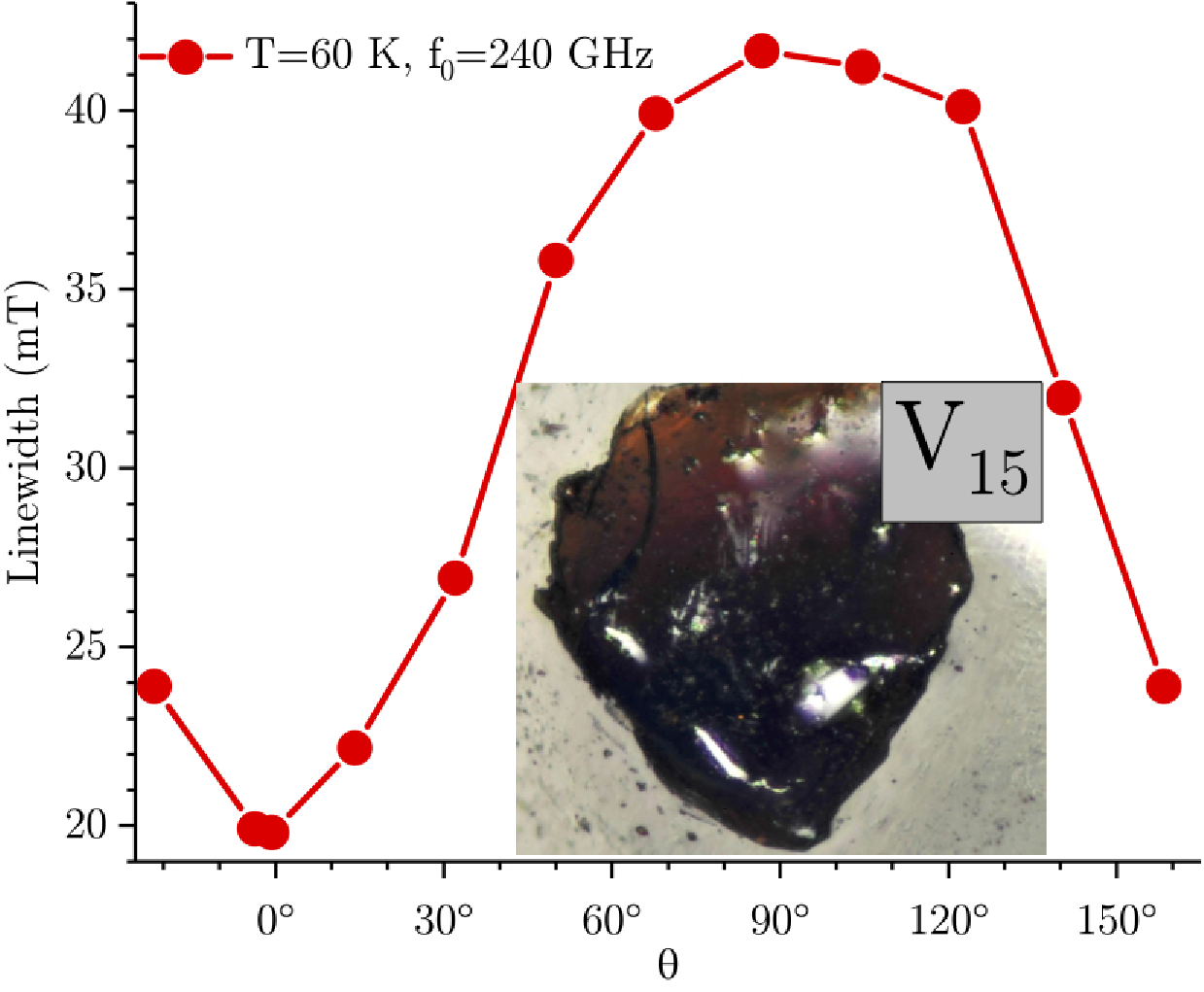}
		\caption{(color online) Experimental linewidth (red dots) measured at 60 K and 240 GHz for a crystal of irregular shape (see insert) as a function of $\theta$. Despite the irregular shape, the linewidth shows the same overal behavior with a minimum/maximum at $\theta=0^\circ/90^\circ$, respectively.}
		\label{figSI_irregular}
\end{figure}

As a consequence of the long range dipolar interactions, each molecule will see a local field summing the individual dipolar fields generated by the other molecules. Therefore, the local field has a certain total value and also a standard deviation due to the spread in values of the individual fields. By means of numerical simulation, one can analyse the dependence on $\theta$ of the spread in dipolar fields and compare it to the observed angular dependence of the experimental resonance linewidths. As a reminder, the main observation in the case of V$_{15}$ is that the linewidth is small/large when the $g$-factor is large ($\theta=0^\circ$) / small ($\theta=90^\circ$). 

The angular dependence of the distribution of fields is calculated in the middle of an ensemble having 22 molecules on each crystallographic axis, that is a total number of $22\times22\times22=10648$ sites. The size is chosen such that dipolar couplings due to molecules outside of this interaction volume can be neglected. The ensemble is shown in the insert of \autoref{figSI_dipole}. The applied field $B_0$ is large enough to polarize the spins. The total field along $B_0$ is given by $B_{dip}=\sum_q{b_q^{dip}}$ where $b^{dip}_q$ is the individual field generated by a molecule at site $q$ in the middle of the crystal. The result is plotted with a continous red line, while the spread $\sigma^2_{dip}=\sum_q{(b_q^{dip})^2}-(\sum_q{b_q^{dip}})^2$ is shown with a dashed blue line as a function of $\theta$. One note that $\sigma_{dip}$ is significantly smaller than the large linewidths observed experimentally (about an order of magnitude smaller) and also that the simulated angular dependence is not correlated with the measured one. The simulated $B_{dip}$ shows the typical behavior $\propto1-3\cos^2\theta$ and both $B_{dip}$ and $\sigma_{dip}$ do not show the experimentally observed behavior of the linewidth (minimum/maximum values at $\theta=0^\circ/90^\circ$ respectively).

This shows that fluctuations in the geometrical part of $\ham_d$ can not explain the observed angular dependence of the linewidth.

Another way to verify the above statement is to measure a sample of an irregular shape as shown in the insert of \autoref{figSI_irregular}. Given that dipolar interactions are of long range nature, the shape-dependent coefficients $R$ and $r$ could completely change the angular dependence of the linewidth. The data in such case can be analyzed and compared with the case of a well defined crystal, with a shape similar to the insert of \autoref{figSI_dipole} and as the one studied in the main article. The linewidth measured at $T=60$~K and excitation frequency 240~GHz is shown with red dots in \autoref{figSI_irregular}. Despite of an irregular shape of the crystal, the main property of the linewidth is conserved, namely that its angular dependence is opposite to that of the $g$-factor. We are thus compelled to relate the linewidth to fluctuations of terms in $\ham_d$ containing the $g$-factor .

\section{Fitting procedure}

\subsection{Field-frequency conversion}
The measured field linewidths can be converted into frequency units through exact diagonalization of the three-spin Hamiltonian $\ham_{st}$, since the linewidth corresponds to a broadening of its eigenvalues. This is needed since the resonance frequency and field are not in a linear relantionship due to zero field splittings. The method presented below is general and can be applied to any other molecular system. We calculate the minimum and maximum excitation frequencies $f_{min,max}$ leading to resonance fields $B_{max,min}$ using the first moment method~\cite{Martens_PRB_2014}. Their difference is:  
\begin{equation}
\Delta = f_{max}-f_{min}.
\end{equation} 
By measuring $\Delta(0\degree)$ and  $\Delta(90\degree)$, one can construct a system of two equations with two unknowns (see Eq.~\ref{Delta}) to solve for $A$ and $a$. This allows to calculate the curve $\Delta(\theta)$ which is converted back to field units by solving the static Hamiltonian: for two frequencies $f'_{min,max}(\theta)=f_0\pm\Delta(\theta)/2$, we calculate the corresponding resonance fields $B'_{max,min}(\theta)$. The difference $B'_{max}-B'_{min}$ is the fitted Full Width at Half Max (FWHM) as a function of $\theta$. Thus, by using only two data points (linewidths at $\theta=0^\circ$ and $90^\circ$) one can simulate the full angular dependence of the linewidth, for any non-linear relantionship between applied field and resonance frequency. 

\subsection{Frequency and temperature studies}

\begin{figure}
	\centering
	\includegraphics[width=\columnwidth]{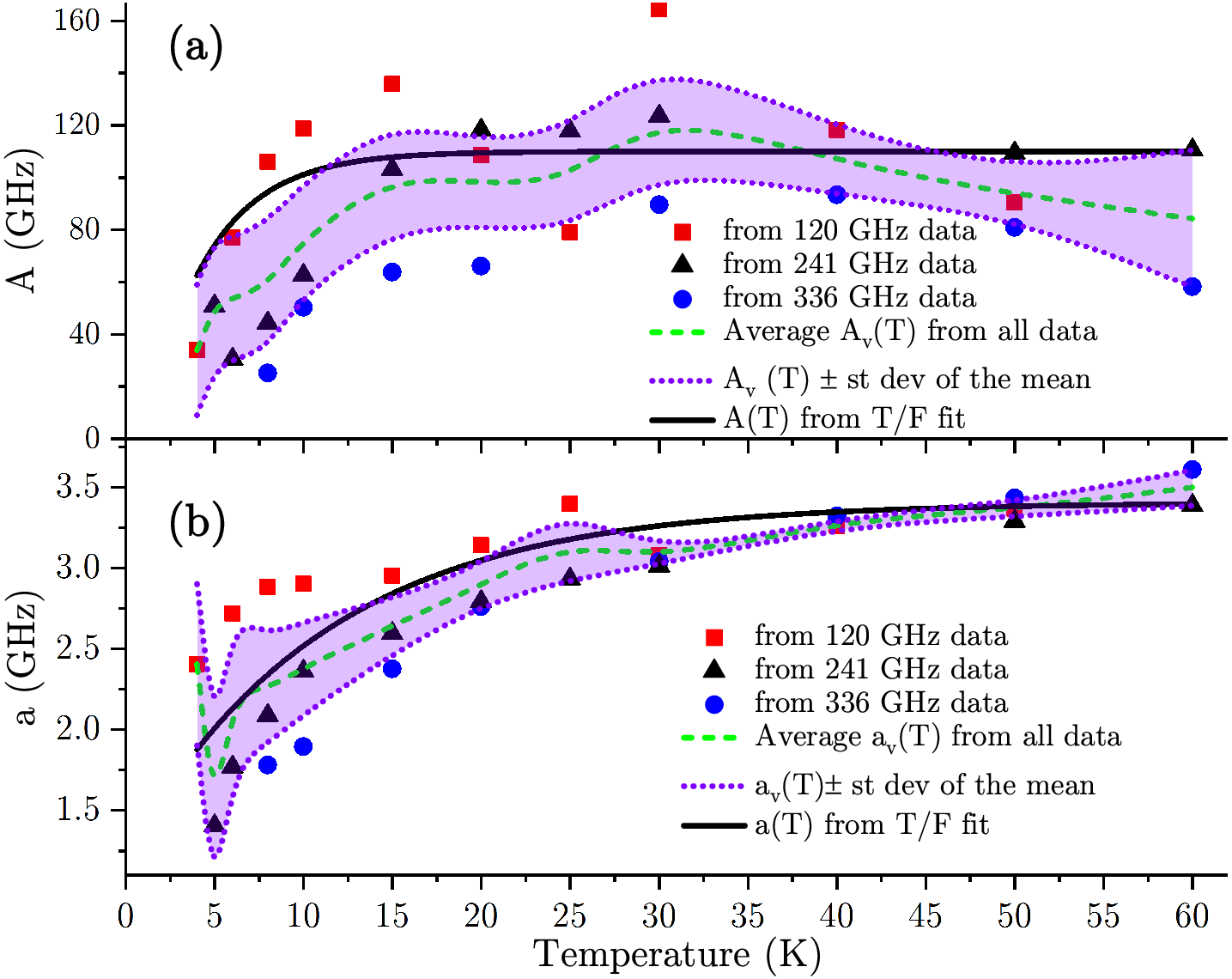}
	\caption{(color online) Fit parameters $A$ and $a$ as a function of temperature, shown in panels \textbf{(a)} and \textbf{(b)} respectively, for three excitation frequencies. Their average is shown with a smoothed dashed green line; two  purple dotted, smoothed lines are bounding the uncertainty area defined as the average $\pm$ the standard deviation of the mean. The black continous line shows an exponential approximation of the temperature trend, used to calculate the temperature / frequency fit (see Fig.~2(c) of the main article).  }
	\label{figSI_Aaparameters}
\end{figure}

The protocol described in the previous section can be applied to find the fit parameters $A$ and $a$ in the temperature interval 4-60~K for the three frequencies used in our study: 120 GHz, 241 GHz and 336 GHz. The obtained values for the fit parameters $A$ and $a$ are shown with scattered dots in \autoref{figSI_Aaparameters}, in panels \textbf{(a)} and \textbf{(b)} respectively: red squares for 120 GHz, black triangles for 241 GHz and blue circles for 336 GHz. The temperature dependence of their averages $A_v(T)$ and $a_v(T)$ is shown with a smoothed dashed green line lying inside an uncertainty area bounded by two  purple dotted, smoothed lines $A_v\pm\sigma_A$ and $a_v\pm\sigma_a$  where $\sigma_{A,a}$ are the standard deviation of the means. 

For both fit parameters one can note an increase with the temperature up $\sim10-20$~K when they reach more stable values. Since $A$ and $a$ reflect the magnitude of the characteristic time $\tau_g$ and $\xi$, one can conclude that the spin-orbit fluctuations are slightly less effective at low temperatures. It is important to take this aspect into account when using our model to fit the T/F master curve shown in Fig.~2(c) of the main article. In \autoref{figSI_Aaparameters}, the black continous lines $A(T)$ and $a(T)$  show exponential approximations of the temperature trends, with characteristic decay constants of 3.6~K and 11~K for $A$ and $a$, respectively. Different decay constants can be linked to different power dependence on $\xi$ for $A$ and $a$, as shown by Eq.~\ref{eq:Aa}. These exponential decays are roughly within the limits of $A$ and $a$ uncertainty bands and are optimized to properly fit the master curves in Fig.~2(c) of the main article (shown with dashed lines).    

\subsection{Discussion of model variables}

The values of parameters $A$ and $a$ can be used to further analyze the variables used by the model given in Eq.~\ref{Deltaxi}. The integrals over the Lorentzian spectra are from 0 to $\Delta$ (see Ref.~[\onlinecite{Bloembergen_PR_1948}]) which leads to:
\begin{align}
a^2&=\frac{\sqrt{2/\pi}}{2\pi}\alpha^2\xi^2R^2\arctan(2\pi\tau_g\Delta) \nonumber\\
A^2&=\frac{\sqrt{2/\pi}}{\pi}\alpha^2\xi^4R^2\arctan(\pi\tau_g\Delta).
\label{eq:Aa}
\end{align}

The $g^2\delta g^2$ term, which defines the value of $a$, is the leading term of the fit procedure and fully describes the angular behavior of the linewidth; the $\delta g^4$ term provides only a marginal improvement of the fit. Consequently, the parameter $A$ is particularly susceptible to a large uncertainty due to the small value of the $\delta g^4$ term: in order to have a meaningful contribution to the fitting procedure, $A$ has to be quite large. Therefore, taking the ratio $A/a$ as an estimation of $\xi$ can lead to unrealistic values for the size of the $g$-factor fluctuations. 

Instead, one can use $a$ to asses the value of $\xi$, after evaluating $R$. As a reminder, the coefficient $R$ is an average of $\sum_{p\neq q}R_{pq}$ (as defined in the main article) and it depends on the spin density and sample shape. An estimation of $R$ can be made by fixing one site, as in the simulations presented in Section III B, that is $R=N\sum_{q}R_{pq}$ with $p$ fixed in the center of the crystal and $N=10,648$ is the total number of sites in the interaction volume. The dipolar field $B_{dip}$, shown in Fig.~\ref{figSI_dipole} with a red continous line, can be estimated as $B_{dip}\sim \frac{\mu_0}{4\pi} g\mu_B S R/N$, which for $g=2$, $S=3/2$ and $B_{dip}\approx 0.5$~mT, leads to $R\sim2000$~nm$^{-3}$ .

Since the linewidth of the homogenous ESR signal in V$_{15}$ is quite large, compared to other spin systems proposed  as qubits, one can assume that $\tau_g\Delta>1$, which is sufficiently large to approximate the $\arctan$ function with $\pi/2$ and thus $a\sim\alpha\xi R$. Assuming a value of $a$ of $\sim 2$~GHz, one get an estimation of the order of magnitude to be $\xi R\sim 100$~nm$^{-3}$ and $\xi\sim0.01-0.1$ as an order of magnitude.  \\

\end{document}